\documentclass[acmsmall, nonacm]{acmart}
\AtBeginDocument{%
  }

\setcopyright{acmlicensed}
\copyrightyear{2025}
\acmYear{2025}
\acmDOI{XXXXXXX.XXXXXXX}

\acmJournal{JACM}
\acmVolume{37}
\acmNumber{4}
\acmArticle{111}
\acmMonth{8}

\usepackage{paralist}

\usepackage{booktabs}
\usepackage{array}
\usepackage{xspace}

\newcommand{\eg}{\textit{e.g.,}\xspace}

\begin{document}

\title{Interaction, Process, Infrastructure: \\A Unified Framework for Human–Agent Collaboration}

\author{Yun Wang}
\email{wangyun@microsoft.com}
\author{Yan Lu}
\email{yanlu@microsoft.com}
\affiliation{%
  \institution{Microsoft Research Asia}
  \country{China}
}








\renewcommand{\shortauthors}{Wang et al.}

\begin{abstract}

While AI tools are increasingly prevalent in knowledge work, they remain fragmented, lacking the architectural foundation for sustained, adaptive collaboration. We argue this limitation stems from their inability to represent and manage the structure of collaborative work.
To bridge this gap, we propose a layered conceptual framework for human-agent systems that integrates Interaction, Process, and Infrastructure. Crucially, our framework elevates Process to a first-class concern, an explicit, inspectable structural representation of activities. The central theoretical construct is Structural Adaptation, enabling the process to dynamically reorganize itself in response to evolving goals. We introduce a five-module Process Model as the representational basis for this adaptation.
This model offers a unified theoretical grounding, reimagining human-agent collaboration as a coherent system for complex real-world work.
\end{abstract}

\maketitle


\section{Introduction}
Today's AI landscape is populated by powerful yet isolated tools. Language models craft polished text, coding assistants generate sophisticated algorithms, and specialized agents automate complex tasks. While each tool demonstrates impressive capabilities individually, these systems exist largely in isolation, disconnected from each other and unaware of their broader context, creating significant practical challenges. Users regularly switch between intelligent systems. A writing assistant may generate a project plan, but has no awareness of the code later written to implement it. A research summarizer may extract key findings, but has no sense of how they are interpreted in a broader analysis pipeline. This result is predictable: users act as \textit{human glue}, repeatedly restating intent, manually tracking dependencies, and stitching together disparate tools.

Recent research begins to explore parts of this puzzle through systems that exhibit agentic behavior. Integration frameworks \cite{qin2024tool, patil2024gorilla, liang2024taskmatrix, shen2023hugginggpt} connect foundation models with external tools and APIs. Autonomous systems \cite{autogpt2023, langchain2022} enable multi-step planning and execution; Adaptive and generative UIs surface contextual actions~\cite{beaudouin2021generative}; while protocols like Model Context Protocol coordinate interactions across applications. Yet most of the techniques focus on AI-to-AI or AI-to-tool orchestration, sidelining the human. Users are either asked to predefine rigid workflows or left watching autonomous executions unfold, functioning more as \textit{disposable utilities} than \textit{collaborative partners}, requiring humans to be equipped with \textit{prompt-engineering} skills just to gain meaningful access.

In contrast, real-world professional workflows (\eg software engineering, scientific research, or creative design) rarely follow predefined steps. Goals shift, constraints emerge, and strategies evolve. Professionals explore, reflect, revise, and adapt. They uncover the problem itself through the act of working. Effective human-agent collaboration should embrace this complexity, supporting human judgment and shared adaptation.

This motivates our central argument: what is missing is not AI capability, but the systems' inability to represent and adapt the structure of collaborative work. Current systems treat collaboration as a sequence of tasks, conversations, or execution traces, rather than as an evolving activity with changing interpretations, goals, workflows, and dependencies. Without an explicit and adaptable representation of process, systems cannot maintain continuity alongside the human. We call this missing capability structural adaptation.

This paper offers a conceptual and theoretical framework for representing collaborative activity and its evolution. Our goal is to articulate the representational foundation needed for AI-native collaboration. We contribute:
(1) \textbf{Structural adaptation as a core theoretical construct}.
We identify and formalize structural adaptation, the capacity of a collaborative process to reorganize its own structure as goals, constraints, and interpretations change. We characterize how this construct explains the structural limitations of current AI systems.
(2) \textbf{A process-based representational foundation for collaborative activity.}
We introduce process as the explicit representation that supports structural adaptation. We define the representational requirements of collaborative activity and propose a minimal, domain-agnostic five-part process model that captures problems, workflows, operations, environment, and reflection.
(3) \textbf{A conceptual framework for process-first collaboration.}
We outline a three-layer conceptual framework that situates process within system design and illustrates how explicit, persistent process representations can support long-term collaboration.
Together, these contributions provide a unified theoretical foundation for process-first, structurally adaptive human-agent collaboration.




\section{Related Work}
Research on human–agent collaboration intersects three major areas: mixed-initiative interaction, activity-centric theories, and agent or workflow architectures. While each line of work contributes important insights, none a mechanism for evolving that structure as human and agent jointly refine their understanding of the work. We outline most relevant traditions and their limitations with respect to this need.

Mixed-initiative (MI) frameworks examine how humans and automated services negotiate control, delegate actions, and coordinate initiative~\cite{horvitz1999principles, allen1999mixed}. Classic MI systems introduced principles for adjustable autonomy, forming the backbone of intelligent user interfaces. However, MI models assume the structure of the task is fixed or implicit, and do not represent goals and workflows as editable objects. Recent research on human–AI collaboration similarly emphasize autonomy, transparency, and role negotiation \cite{shneiderman2022human, seeber2020machines,li2024we}, but mainly focus on interactional or social dynamics.
Our work addresses a different layer of collaboration: explicit, semantic representations of activity structure and their adaptation over time. A process-first perspective exposes the structure itself. This capability of structural adaptation is essential for open-ended human-agent collaboration.

Activity theory \cite{engestrom2001expansive}, distributed cognition \cite{hutchins2000distributed,Suchman1987}, and sensemaking frameworks \cite{pirolli2005sensemaking, russell1993cost} describe how humans structure and evolve their own activities through interpretation and mediation, offering strong conceptual grounding. They emphasize that activity, not task, is the meaningful unit of human work. Activity-based computing further operationalizes by treating activities as first-class computational constructs across sessions \cite{bardram2005activity,Bardram2006}. However, activity-based theories and systems model human behaviors, analyze how people coordinate with tools or with each other, and treat tools as passive resources. Our work extends these foundations by introducing an explicit representation of collaborative process, with structure can be reshaped as human and agent co-develop the activity. It enables agentic partners capable of modifying goals, workflows, or representations. 

Recent LLM-based agents can generate multistep plans, invoke tools, and perform iterative refinement \cite{Yao2023ReAct, shinn2023reflexion, yao2023tree}. Workflow and planning systems similarly encode task sequences and control flow through explicit schemas \cite{Lacity2016, van2022process, OMG2011BPMN}, but they are fundamentally execution-centered: plans are constructed internally, and workflows remain difficult for users to inspect or modify. In addition, recent coordination frameworks such as  Model Context Protocol (MCP) focus on agent-to-agent interoperability, but they do not provide a manipulable representation of the collaborative activity that can evolve with human intent. Our work focuses on how the structure of the collaborative activity is represented and adapted over time, offering capabilities essential for open-ended human–agent collaboration.

\section{Collaborative Activity, Process, and Structural Adaptation}\label{sec:construct}
We characterize collaborative activity as the unit of work, introduce process as its structural representation, and develop structural adaptation as the key capability through which collaborative structure evolves.

\begin{figure}[!htbp]
\centering\includegraphics[width=0.7\linewidth]{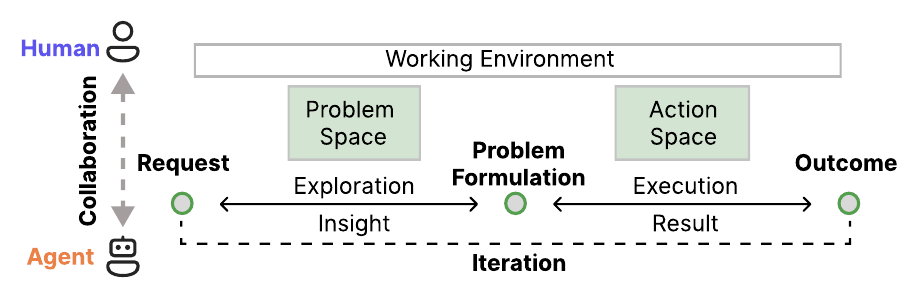}
    \vspace{-5pt}
    \caption{
    Collaborative activity unfolds through iterative cycles in which participants move between the problem space and the action space. Exploration generates insight that reshapes the problem formulation, while execution produces results that inform subsequent steps. This evolving structure motivates the need for an explicit process representation.
    }
    \vspace{-5pt}
    \label{fig:collaboration-dynamics}
\end{figure}

\paragraph{Activity as the Unit of Collaborative Work}
Collaborative work unfolds not as a sequence of isolated tasks but as an evolving activity. An \textit{activity} is a purposeful, meaning-laden unit of work. In real practice, including writing, analysis, design, or research, people navigate cycles of framing, exploration, synthesis, and adjustment. These shifts arise from engagement with the problem as it develops (\autoref{fig:collaboration-dynamics}). 
Activities possess properties essential for collaboration. They are interpretive, requiring ongoing sense-making; structured, involving relationships among goals, constraints, and subproblems; contextual, drawing on heterogeneous information and artifacts; and dynamic, continually evolving as understanding deepens. Systems that respond only to individual commands or task specifications cannot participate in this evolving structure. Effective collaboration therefore requires a representational account of the activity itself.

\paragraph{Process as the Representation of Activity}
The representation of activity's structure is what we call a process. A \textit{process} is a structured depiction of an activity's goal, organization, available operations, contextual dependencies, and current state. Unlike plans, histories, or execution logs, a process is persistent, explicit, and editable. It provides a shared substrate through which both human and AI can interpret what is being done, why it is being done, and how the work is unfolding.
Processes vary in granularity: a process may represent an entire activity or a focused sub-activity. Activities may contain sub-activities, and each can be represented as a process using the same structural schema. What defines a process is not its size but its role as the structured representation of a coherent unit of activity.
Because activities evolve, the process is expected to also evolve, adjusting goals, reorganizing structure, and shifting where work is being advanced. A process enables continuity, shared understanding, and coordinated adaptation over time.

\paragraph{Structural Adaptation as a Core Construct}
As collaborative activities unfold, their structure rarely evolves uniformly. The Components of a process shift at different times, at different rates, and for different reasons. Structural adaptation captures these non-uniform dynamics: it is the capacity of a process to modify its own representational structure as collaboration progresses. This construct explains how human–AI collaboration maintains coherence, alignment, and continuity across extended, evolving work. Structural adaptation takes three fundamental forms. These three forms together constitute the full space of structural change necessary for activity-level collaboration:
\begin{compactitem}
 \item \textbf{Reframing.}
Changes to the meaning structure of the activity, \eg shifting goals, altering assumptions, rearticulating problem boundaries, or redefining what success entails. Reframing realigns the overarching purpose of the collaborative work.
 \item \textbf{Reorganization.}
Changes to the organizational structure of the process, \eg rearranging steps, restructuring dependencies, introducing new pathways, or consolidating existing ones. Reorganization shapes how work is coordinated and sequenced.
 \item \textbf{Reorientation.}
Changes to where the activity is presently being advanced within the process representation. Reorientation redirecting effort toward a different sub-problem, component, or level of detail, without altering the overall topology of the process. It captures the dynamic movement of the activity's structural frontier.
\end{compactitem}

\section{The Collaborative Process Model}\label{sec:process}
As discussed in \autoref{sec:construct}, without process structure, systems cannot maintain continuity, support reframing, or participate in reorganizing the work. To provide the representational basis for structural adaptation, we introduce a five-module process model that captures the structural elements necessary for representing collaborative work.

\begin{figure}[htbp]
    \centering
    \includegraphics[width=0.9\linewidth]{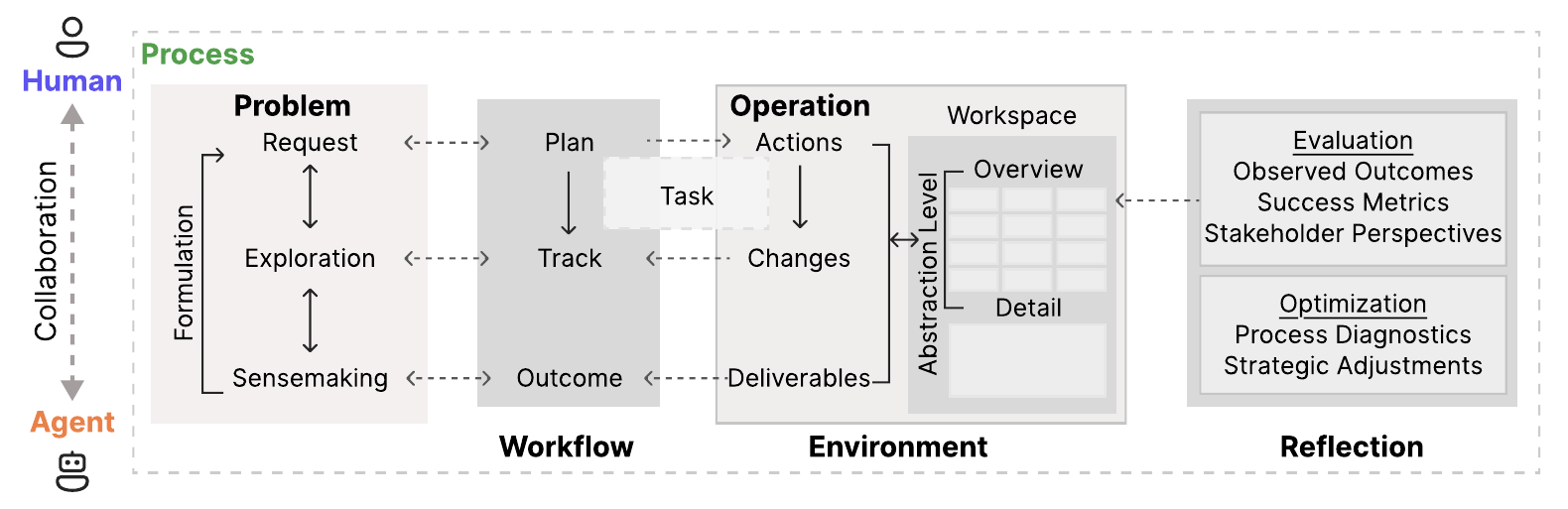}
    \vspace{-5pt}
    \caption{The process as an explicit representation of collaborative activity.
It consists of five structural modules: Problem Space, Workflow, Operations, Environment, and Reflection. They capture the evolving organization, context, and methods of the activity. This representational substrate enables structural adaptation across reframing, reorganization, and reinterpretation.}
    \vspace{-8pt}
    \label{fig:process}
\end{figure}

\paragraph{The Five Modules}
A process must represent multiple facets of an evolving activity: its purpose, organization, available actions, resources, and the meta-level reasoning through which the work is shaped. The five modules described below constitute a minimal yet sufficient set of representational components. Each module captures a distinct structural dimension, and together they form a coherent substrate for supporting structural adaptation (\autoref{fig:process}).

\textbf{Problem Space} The Problem Space represents the evolving interpretation of the activity, its goals, assumptions, hypotheses, constraints, and framing. It captures what the work is about and how the problem is currently understood. As understanding deepens or shifts, the problem space can be reframed, allowing the collaboration to redefine aims, criteria, or conceptual boundaries.

\textbf{Workflow} The Workflow part models the organization of the work: its stages, partial orders, dependencies, and parallel threads. It provides a structural scaffold rather than a fixed script. As strategies change, the workflow may be reorganized: reordering steps, merging or splitting paths, or introducing new subactivities. Workflow captures how the work is structured at any moment.

\textbf{Operations} The Operations module specifies the set of actions, procedures, tools, or model capabilities involved during collaboration. It describes what and how tasks are completed. Operations may be high-level reasoning modes, domain-specific methods, tool invocations, or user-defined procedures. As collaboration evolves, operations may be reparameterized or new operations introduced to reflect emerging needs or strategies.

\textbf{Environment}
The Environment is the shared workspace of the collaboration: the artifacts, intermediate results, external resources, available tools, and workspace state on which the work depends. It may include documents, code, datasets, sketches, diagrams, generated outputs, and any tool or resource accessible during the activity. The environment changes dynamically as new artifacts are created, or intermediate results become input for subsequent steps. It provides the material substrate on which operations act and the persistent grounding for continuity over time.

\textbf{Reflection}
The Reflection module encompasses the meta-level processes that examine progress, detect inconsistencies, critique assumptions, and decide when structural change is needed. Reflection coordinates reframing, workflow reorganization, abstraction shifts, and other forms of structural adaptation. It answers why the work should change and how the structure should evolve.

\paragraph{Dynamics and Interactions Among Modules}
Structural adaptation arises from coordinated changes across modules. Reframing in the Problem Space may prompt reorganization in the Workflow; new workflows require updated Operations; newly created artifacts reshape the Environment; and changes in the environment can, in turn, reveal new constraints or opportunities that alter the Problem Space. Reflection mediates these interactions, ensuring that structural changes remain coherent and aligned with the evolving activity. Together, these five modules form a coherent representational scaffold through which human and AI can jointly maintain, inspect, and adapt the structure of collaborative work. These interactions create a feedback loop in which structure is continuously assessed, reshaped, and realigned with evolving human intent.

\section{Framework for Human–Agent Collaboration}
\begin{figure}[htbp]
    \centering
    \vspace{-3pt}
    \includegraphics[width=0.85\linewidth]{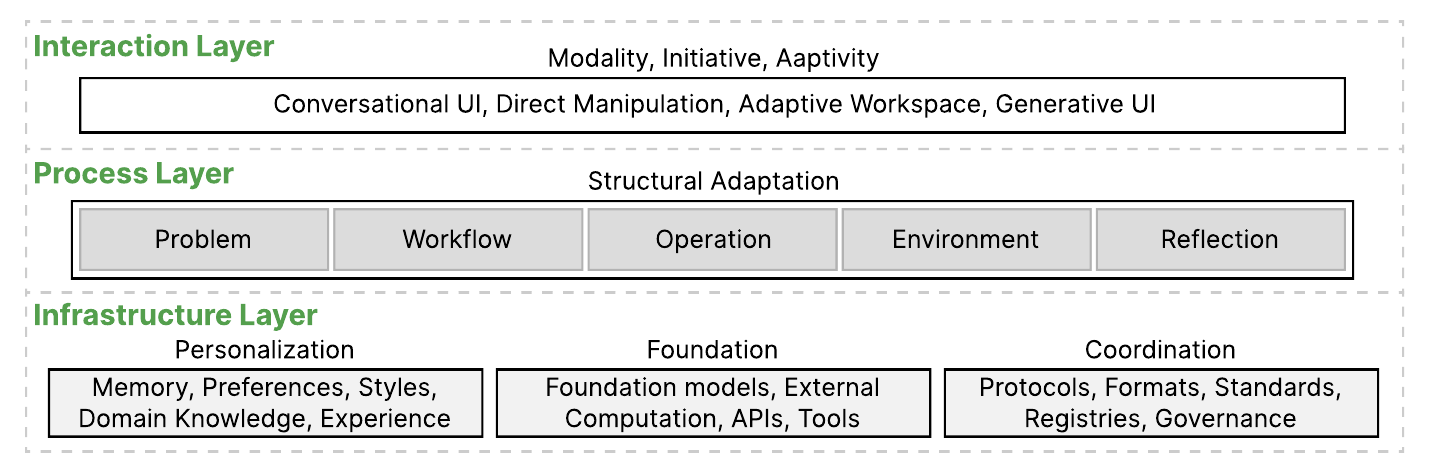}
    \vspace{-5pt}
    \caption{Three-layer conceptual framework for process-first collaboration. The Interaction Layer provides user-facing projections of the activity; the Process Layer maintains the explicit, editable structure of the activity; and the Infrastructure Layer supplies the models, tools, memory, and resources that support execution and continuity.}
    \vspace{-5pt}
    \label{fig:three-layer}
\end{figure}

With Process as the representational substrate that links user intent with system behavior, we propose a unified architecture that separates interaction, representation, and execution into three conceptual layers. This framework treats collaboration as a system-level capability supported by explicit process representation (\autoref{fig:three-layer}): 
(1) an Interaction Layer, where users express intent and inspect or modify structure;
(2) a Process Layer, which maintains the explicit, evolving representation introduced; and
(3) an Infrastructure Layer, which grounds operations in model capabilities, external tools, and computational resources.
The framework clarifies how structural adaptation is exposed to users, how it is maintained over time, and how it is operationalized through available tools and resources. In doing so, it frames collaboration as a representational capability rather than a sequence of tasks, enabling both humans and AI systems to participate in shaping the work.

\paragraph{Interaction Layer}
The Interaction Layer provides the projections through which users encounter and shape the process. This layer exposes structured representations of the activity, allowing users to interpret the work and influence its evolution. Interactions vary along two fundamental dimensions. \textit{Modality} determines how structure is expressed (\eg text, voice, sketches, diagrams, canvases, etc.). \textit{Initiative and control} determine how collaboration unfolds, whether user-driven, system-suggested, or mixed-initiative. These dimensions shape how users participate in the collaboration: articulating intent, refining assumptions, steering direction, or restructuring parts of the process.

A central property of this layer is decoupling user interface from process logic. The same underlying process may be projected as a conversational exchange, a workflow diagram, a timeline, or a spatial workspace, depending on user preference, task context, or stage of work. These projections provide different views of process representation, enabling structural inspection and manipulation at appropriate levels of abstraction. This decoupling echoes prior work on instrumental interaction~\cite{beaudouin2000instrumental}. However, in our context, instruments serve beyond mediators of action, but the evolving process that reveals different stages of work. The layer supports more than input and output. Users can interrogate decisions, examine dependencies, annotate assumptions, and request reframing or reorganization. In this sense, the interaction layer aligns user interactions with underlying activity.

\paragraph{Process Layer}
The Process Layer serves as the representational core of the architecture. It maintains the explicit, evolving structure of the collaborative activity as introduced in \autoref{sec:process}. These modules form the semantic substrate through which collaboration is interpreted, guided, and adapted. The Process Layer preserves continuity across time, tools, and collaborators and records the rationale behind decisions and maintains the representational state that makes different forms of collaboration possible.

Because it sits between interaction and execution, the Process Layer also mediates cross-layer propagation. User actions at the Interaction Layer, such as reframing goals, restructuring workflows, annotating assumptions, or selecting operations, are interpreted as modifications to the process representation. Conversely, updates within the Process Layer guide the Infrastructure Layer by determining which operations are valid, which tools should be invoked, and how results should be integrated or interpreted. This positioning establishes a stable locus of meaning across the system: a shared, inspectable representation that both humans and AI systems rely on to coordinate action, maintain alignment, and evolve the activity over time.

\paragraph{Infrastructure Layer}
The Infrastructure Layer provides the computational substrate that grounds the process in executable capabilities. It operationalizes the structures maintained in the Process Layer. While the Process Layer specifies what the activity is and how it is structured, the Infrastructure Layer determines how operations are carried out and how their results return to update the representation. 
This layer may integrate a wide range of mechanisms, including foundation models, domain-specific APIs and tools, contextual memory systems, and communication protocols. These components supply the computational resources required to enact the operations. Because processes evolve, the infrastructure must support flexible, heterogeneous, and extensible forms of execution, allowing different models, tools, or data sources to be combined or substituted without altering the representational state.

A key property of this layer is its decoupling from how processes are represented or projected. The same process structure can be enacted through different computational pathways depending on available tools, model capabilities, or task constraints. This separation enables scalable, multi-tool collaboration while preserving a stable semantic core in the Process Layer. It ensures that structural adaptations translate coherently into action, while execution details remain modular and replaceable. Through this coordination between representation and execution, the Infrastructure Layer completes the architectural stack, enabling adaptive collaboration to be realized in real tasks and environments.
%

\section{Representational Analysis of AI Systems}\label{sec:evaluation}
\begin{figure}[!htbp]
    \vspace{-3pt}
    \centering\includegraphics[width=.8\linewidth]{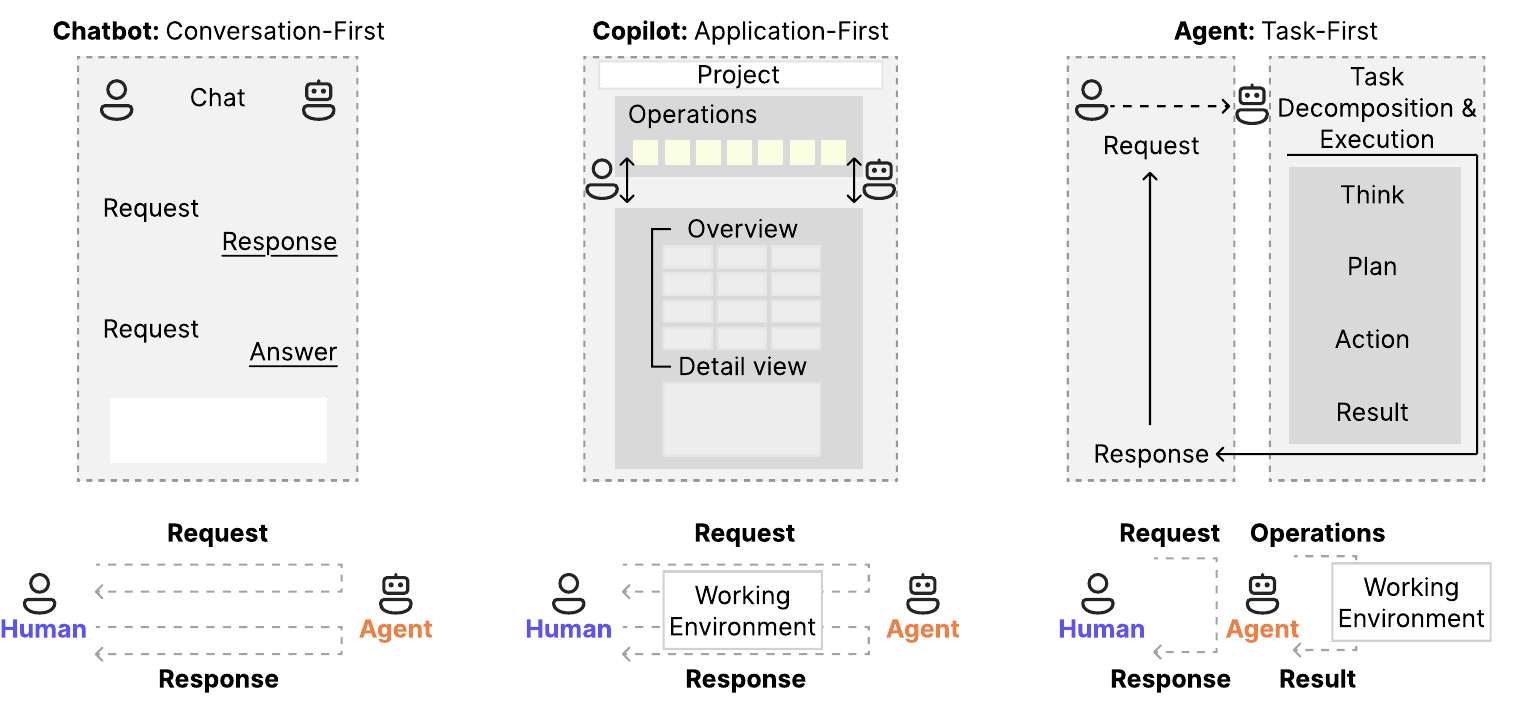}
    \vspace{-5pt}
    \caption{Representative forms of systems in today's AI landscape: chatbots, copilots, and agents. Each encodes a distinct model of collaboration but remains isolated, lacking connective logic across tasks and interfaces.}
    \label{fig:compare}
    \vspace{-5pt}
\end{figure}

Contemporary AI systems commonly appear as conversational assistants, application-embedded copilots, or task-oriented agents (\autoref{fig:compare}). Although these systems operate differently, they share a common characteristic: each encodes implicit assumptions about how collaborative work is structured. 
We do not aim to provide a taxonomy, nor an exhaustive categorization. 
Instead, this section applies the representational model as an analytical lens to illustrate how contemporary system forms express, or fail to express the structural elements required for collaborative activity, demonstrating the explanatory power and diagnostic clarity of the proposed framework.

\paragraph{Structural Adaptation Lens}
Viewed through the lens of structural adaptation, all three system forms exhibit the same limitation: they treat changes in activity implicitly, locally, and reactively, making it difficult for users and systems to jointly evolve the work.

\textbf{Chat-based assistants} Chat interfaces support a fluid exchange of language but provide no representational structure for the activity. All history is compressed into a linear conversation. Users can request new directions, but each redirection is handled as a local conversational turn rather than as a modification of an underlying activity structure. 

\textbf{Application-embedded copilots} Copilots offer contextual assistance anchored to a specific application, enabling local adjustments but not structural change.
They can refine code within an editor, rewrite text within a document, or summarize a dataset inside a notebook, yet the underlying activity remains outside the system’s representational horizon. Any shift in direction requires users to manually realign artifacts across tools.

\textbf{Task-oriented agents} Agents often possess internal plans or tool-use sequences, but these plans are opaque, brittle, and difficult for users to inspect or reshape. Once execution begins, deviations from the anticipated structure cause plans to fail or restart. Adaptation occurs only within the narrow confines of a predefined task.

\paragraph{Process Analysis}
The five components of process provide a representational vocabulary for describing the structure of a collaborative activity.
Applying these components to existing AI system forms makes visible how each system organizes activity structure and which structural elements remain implicit, inaccessible, or absent (\autoref{tab:process-evaluation}).
\begin{compactitem}
\item \textbf{Problem Space Representation} 
Chat-based assistants treat the problem space as implicit in the conversation.
Goals must be continually restated and cannot be maintained as a coherent representation.
Copilots ground the problem space in a specific application.
Agents typically generate an internal problem framing when initiating a plan, yet this framing usually remains opaque and cannot be reshaped or negotiated by the user. 
\item \textbf{Workflow and Structural Rigidity}
Chat interfaces provide no workflow representation: all progress collapses into a linear turn-by-turn trace.
Copilots operate within the workflow implicitly built into the host application, reflecting the software's operational model, such as editing a file, modifying code cells, and are fixed by design. Because these structures are predefined, they offer no representation of the activity’s broader workflow.
Agents construct workflows by decomposing a user-specified goal into an internal plan. This plan guides tool calls and action sequences, but it cannot be reshaped, extended, or reorganized once generated.
\item \textbf{Operation Space Exposure}
Chat-based systems do not expose a structured set of operations. Interactions take the form of open-ended language requests.
Copilots surface operations that are bound to the application. Agents leverage tool APIs and planners, yet the available operations and their sequencing remain inaccessible to users, limiting opportunities for mixed-initiative organization.
\item \textbf{Environment as Shared Context}
In chat-based systems, the environment is confined to the transient context window. Copilots access the local environment of the application but cannot integrate information across applications. Agents maintain internal memory, but these remain unmanageable to users for collaboration.
\item \textbf{Reflection and Process-Level Feedback}
Across all three system forms, reflective activity remains largely limited.
Chat-based assistants do not maintain a representation of what has been learned, reconsidered, or revised.
Copilots may offer suggestions or alternative edits, yet these moments arise from local heuristics rather than from a process that the user can inspect or participate in.
Agents sometimes perform internal retries or self-corrections. But these reflective adjustments cannot coordinate with user-led reframing or restructuring.
\end{compactitem}

\begin{table}[!htbp]
\centering
\small
\begin{tabular}{p{2.5cm} p{2.8cm} p{2.8cm} p{3cm}}
\toprule
\textbf{Process Module} & \textbf{Chat} & \textbf{Copilot} & \textbf{Agent} \\
\midrule
\textbf{Problem Space} & Implicit & Local & Internal \\
\textbf{Workflow} & None & software-implied & Fixed-plan structure  \\
\textbf{Operation} & Language-driven & software-defined & Planner-driven, hidden \\
\textbf{Environment} & Context window & Local workspace & Internal state \\
\textbf{Reflection} & None & Local suggestions & Opaque adjustments \\
\bottomrule
\end{tabular}
\caption{Process modules expressed in three contemporary AI system forms.}    \label{tab:process-evaluation}
\vspace{-5pt}
\end{table}
\vspace{-10pt}

\paragraph{Architectural Analysis}
The three-layer architecture offers a system-level view on how contemporary AI systems organize interaction, reasoning, and execution. Examining through this perspective reveals how distinct architectural commitments shape what users can see, modify, or coordinate during collaborative work.

\textbf{Interaction Layer}
Chat-based assistants expose a single interaction surface: a conversational box.
Intent, clarification, and feedback all pass through text exchanges, making interaction flexible but limiting.
Copilots ground interaction in the host application's interface. Users operate through familiar controls, and the AI's suggestions are tightly coupled to the artifact at hand.
Agent systems vary widely. Some rely on text input with a chat-like surface; others expose simple task forms or parameter panels. 
Automation-oriented platforms, such as RPA tools, present node–link flows that show what the agent is doing \cite{willcocks2016robotic}. These visualizations enable definition of automation sequences rather than editable representations of an evolving activity.

\textbf{Process Layer}
Across these system forms, the process layer is either absent or implicit.
Chat systems maintain no explicit representation of goals, workflow, or reflective state.
Copilots follow the software's built-in operational logics.
Agents generate internal plans or states.
None of these systems offer a shared process representation that could support structural adaptation.

\textbf{Infrastructure Layer}
The infrastructure layer is where current systems differ most.
Chat-based assistants rely primarily on model inference.
Copilots combine model inference with application APIs, enabling richer operational grounding but keeping it local to the tool.
Agents integrate planners, retrieval modules, and tool APIs, forming complex execution pipelines. They allow the agent to perform operations, but not to act on process representations or expose the intermediate structure of the activity.
Viewed through the architectural lens, current AI systems differ in how they distribute capability across interaction and infrastructure layers, yet share a common absence in the process layer.
This absence shapes how collaboration unfolds: users interact with surfaces or execution traces rather than with the evolving structure of the activity.
As a result, adaptation remains local to the system’s interaction modality or execution pipeline, rather than grounded in a shared model of the activity.

\paragraph{Synthesis}
The three analytical lenses—structural adaptation, process components, and architectural layering—converge on a consistent pattern: Contemporary AI systems support local task execution but do not maintain an explicit representation of the evolving activity.
Across the five components of process, many elements of activity remain implicit, fragmented, or inaccessible.
Viewed architecturally, these limitations reflect the absence of a process layer.
Taken together, these analyses demonstrate how the representational model introduced in this paper can be used to examine contemporary systems.
Here we apply the model to three broad system forms, but the same lenses can be used at finer granularity to analyze individual systems or components.

\section{Research Challenges and Opportunities}\label{sec:opportunities}
The layered model presented in this paper recasts collaborative intelligence as a system-level human-agent interaction and coordination. 
Effective support for open-ended, evolving work demands explicit representations of process across time, tools, and participants. 

\paragraph{Toward a Unified, Adaptive Framework}
As human-agent collaboration becomes more pervasive, the lack of a unified representation of work remains a fundamental bottleneck. A unified architecture offers structural coherence and design flexibility. Like the von Neumann model in computing~\cite{vonNeumann1945} or the MVC pattern in UI design~\cite{Krasner1988}, our model treats collaboration as modular and composable. It enables reuse of components (\eg planning, memory) and extension of capabilities without rearchitecting the entire stack. It also supports \textit{epistemic generalization}: while tools vary, domains such as design, law, and research often share high-level process patterns.

As analyzed in \autoref{sec:evaluation}, Current AI tools are largely static. Adaptation, when it occurs, is often local and reactive. Our framework addresses this by combining modular separation with explicit linkage. Each layer is independently evolvable yet structurally aligned, allowing updates in one layer to propagate to others.  
More importantly, because the process layer is explicitly modeled, the system gains visibility into task structure, delegation patterns, and breakdowns. This enables the capture of structured traces of collaboration, data that can support meta-reasoning across the human-agent workflow. Future agent system can then revise workflows, adjust interface patterns, reallocate tools, or propose model changes. In this way, ongoing experience fuels \textit{architectural self-improvement}.

\paragraph{Interfaces that Adapt to Users, Scenarios, and Evolving Processes}

Interaction in AI systems has largely focused on usability, responsiveness, and intuitive input/output modalities. Process-aligned systems require a broader perspective: interfaces are not merely endpoints for command issuance or response display, but \textit{active surfaces} for perceiving, guiding, and reflecting on collaborative workflows. 
This perspective draws on foundational HCI concepts such as \textit{interface transparency}~\cite{Norman1988}, \textit{mixed-initiative coordination}~\cite{horvitz1999principles}, and \textit{calm computing}~\cite{Weiser1996}, while extending them toward dynamic, process-aware interaction. 
To support such workflows, interfaces face three interrelated challenges: 
(1) \textit{Dynamic Representation.} Interfaces need to adapt process granularity (overview vs.\ detail), modality (text, diagrams, timelines), and semantic framing (goal-oriented vs.\ action-oriented) to fit evolving contexts. Designing these representational grammars is nontrivial. Prior work on adaptive interfaces~\cite{Shneiderman1998} and progressive disclosure~\cite{Norman1988} provides useful principles, yet process-aligned systems call for richer models.
(2) \textit{Cross-Level Navigation.} Professionals frequently shift across abstraction levels. Interfaces should support bidirectional navigation: from plans to operations, annotations to execution traces, or breakdowns to causal rationale. This recalls semantic zooming~\cite{perlin1993pad}, but with an emphasis on aligning mental models across users, agents, and shifting process state.
(3) \textit{Attention Orchestration.} Managing user attention becomes essential as workflows evolve. Systems need ways to surface process drift, highlight misalignments, or prompt reflection without overwhelming users. This challenge revisits classic notions of system visibility~\cite{Norman1988} and calm technology~\cite{Weiser1996}, now situated in dynamic, multi-stage workflows. Design questions arise: When should the system indicate ambiguity? How might it encourage recalibration? What cues best elicit user feedback on process plans?

\paragraph{Co-Evolution and Longitudinal Human-Agent Collaboration}
Beyond short-term performance or flexibility, process-aligned systems open the door to longitudinal co-evolution between humans and agents. Rather than acting as disposable utilities or context-free assistants, future agents could develop shared memory, mutual understanding, and evolving collaboration patterns that persist across sessions, tools, and goals.
Evaluating such dynamics requires methodological innovation. More importantly, co-evolution is not one-sided. As systems adapt, so do users. Agents should not only respond to input, but also facilitate meta-cognition, helping both parties reflect on \emph{how} they are collaborating and \emph{how} that collaboration is changing. This reframes the agent's role from reactive assistant to reflective partner, capable of maturing alongside its user, adjusting to evolving contexts, and sustaining the logic of complex, unfolding work.

\paragraph{Limitations and Tradeoffs}
Explicit process representations introduce several limitations and tradeoffs. First, maintaining structured activity representations may create representation overhead, and not all tasks justify this additional cognitive and computational cost. Second, explicit structure risks over-structuring early, open-ended, or exploratory work, potentially biasing how problems are framed even though structural adaptation enables later revision. Third, collaborative restructuring introduces coordination costs: inconsistencies, invalid dependencies, or divergent interpretations between human and AI may require additional effort to detect and resolve. Fourth, making process structure visible improves transparency but does not eliminate trust tensions, as users may still question the validity of AI-proposed structural changes.
Finally, the framework presumes that operations from heterogeneous models and tools can be meaningfully organized under a unified process representation—an aspiration that may be difficult to realize in practice. These limitations do not undermine the conceptual contribution but instead clarify the boundaries of the framework and point to open challenges for future process-first systems.
\section{Conclusion}
The rise of increasingly capable AI systems has unlocked new possibilities for supporting human creativity, analysis, and decision-making. Looking ahead, we advocate a shift from task augmentation to process-first system design. Our layered architecture, spanning interaction, process, and infrastructure, offers a conceptual foundation for building intelligent systems that are not only capable but coordinated; not only responsive but reflective. By embedding shared logic at the core of human-agent systems, we open the door to a new generation of tools that can adapt, evolve, and genuinely collaborate with their users.

\bibliographystyle{ACM-Reference-Format}
\bibliography{main}










\end{document}